\documentclass[journal=jpclcd,manuscript=article]{achemso}

\usepackage{chemformula} 
\usepackage{tabularx}
\usepackage{graphicx}

\begin{tocentry} 
\includegraphics[width=2in]{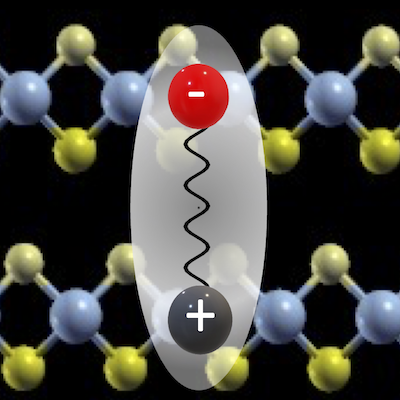} 
A schematic figure of interlayer excitons in Janus bilayers.
\end{tocentry}
\author{F. Nilsson}
\email{fanni@dtu.dk}
\affiliation{CAMD, Department of Physics, Technical University of Denmark, 2800 Kgs. Lyngby, Denmark}
\alsoaffiliation[2]{CPHT, CNRS, Ecole Polytechnique, Institut Polytechnique de Paris (l'X), F-91128 Palaiseau, France}

\author{M. Kuisma}
\affiliation{CAMD, Department of Physics, Technical University of Denmark, 2800 Kgs. Lyngby, Denmark}
\author{S. Pakdel}
\affiliation{CAMD, Department of Physics, Technical University of Denmark, 2800 Kgs. Lyngby, Denmark}
\author{K. S. Thygesen}
\affiliation{CAMD, Department of Physics, Technical University of Denmark, 2800 Kgs. Lyngby, Denmark}

\title[Bilayer]{Excitonic Insulators and Superfluidity in 2D Bilayers Without External Fields}

\begin{document}

\begin{abstract}
We explore a new platform for realising excitonic insulators, namely van der Waals (vdW) bilayers comprising 2D Janus materials.
In previous studies, Type-II heterobilayers have been brought to the excitonic insulating regime by tuning the band alignment using external gates. In contrast, the Janus bilayers of the current work represent \emph{intrinsic} excitonic insulators. We first conduct \emph{ab initio} calculations to obtain the quasiparticle band structures, screened Coulomb interaction, and interlayer exciton binding energies of the bilayers. These \emph{ab initio}-derived quantities are then used to construct a BCS-like Hamiltonian of the exciton condensate. By solving the mean-field gap equation, we identify 16 vdW Janus bilayers with insulating ground states and superfluid properties. Our calculations expose a new class of advanced materials that are likely to exhibit novel excitonic phases at low temperatures, and highlight the subtle competition between interlayer hybridization, spin-orbit coupling, and dielectric screening that govern their properties. 
\end{abstract}


Excitonic insulators can be formed in semiconductors or semimetals with sufficiently large electron-hole interactions such that the exciton binding energy dwarfs the band gap energy. Excitons are bosonic quasiparticles formed by bound electron-hole pairs and can therefore condensate as well as exhibit superfluidity at sufficiently low temperatures. Although proposed already in the 1960s\cite{mott1961,halperin1968}, it was not until very recently that experimental evidence of excitonic insulators has been found.  
Experimental indications of exciton condensation in 1T-TiSe$_2$ was reported 2017 \cite{kogar2017} and in monolayer WSe$_2$ \cite{jia2022} and WTe$_2$ \cite{sun2022} as well as bulk Cu$_2$O \cite{morita2022} this year.  

In bulk and monolayer materials with indirect bandgaps, such as 1T-TiSe$_2$, the exciton formation is typically accompanied by a lattice instability or charge-density wave (CDW). The exciton-driven CDW can be difficult to disentangle from the corresponding  instabilities driven by other mechanisms
and also tends to inhibit superfluidity\cite{kogar2017}. 
It was recently shown that even condensation of zero-momentum excitons (formed in materials with direct band gap) can drive a CDW instability due to the hybridization between the conduction and valence bands, which in general is non-negligable for bulk materials.\cite{Mazza2020}
In bilayer materials, on the other hand, the electrons and holes can be separated in real-space in the two layers. Due to the spatial separation between the conduction and valence band both the hybridization and the electron-phonon coupling is greatly reduced, thus reducing the probability for lattice instabilities\cite{gupta2020}. 
Experimental evidence of exciton condensation in bilayer materials was first reported for materials in high magnetic fields\cite{liu2017,li2017} but later also  in double bilayer graphene\cite{Burg2018} and very recently an exciton insulating phase was reported in WSe$_2$-MoSe$_2$ bilayers \cite{ma2021}. 
Common to all these bilayer architectures is that the electron and hole populations are controlled by external gates and that the two layers are separated by a thin dielectric barrier, typically chosen to be hexagonal boron-nitride (hBN). While this tunability allows for a flexible device that can be easily controlled, the insulating barrier also decreases the electron-hole interactions which, in turn, yields reduced exciton binding energies and consequently smaller superfluid transition temperatures. In this article we propose an alternative strategy to achieve equilibrium excitonic insulators in bilayer architectures without insulating barriers or external gates. 

2D Janus materials is an active field of research with novel structures synthesized yearly. Recent examples include few-layer BiTeCl and BiTeBr, which were synthesized 2020 \cite{Zhang17,Hajra20}. In this article we focus on MXY-Janus materials such as MoSSe that have a similar structure as the standard H-phase transition metal dichalcogenides (H-TMDs) but with two different species of chalcogens on the two sides of the 2D sheets. This gives rise to an intrinsic electric field perpendicular to the 2D plane. In Janus bilayers this field will, if stacked with aligned dipoles, substantially decrease the band gap\cite{riis2018efficient}. Thus the intrinsic dipole of these materials play the same role as the external gates in the bilayer devices discussed above. Therefore, by combining different Janus and TMD layers it is possible to engineer the band alignment without the need for external gates and insulating barriers. 

We employ first-principles calculations to predict a number of new excitonic insulators and exciton superfluids with high transition temperatures. The investigated systems are  
vdW bilayers composed of already synthesized monolayers as well as stable, but not yet synthesised monolayers from the Computational 2D Materials Database (C2DB)  \cite{c2db1,c2db2,c2db3}. In particular, we show that homobilayers of Janus materials can become excitonic insulators with potential to condensate and exhibit superfluidity at low temperatures. These conclusions are reached by solving a BCS-like gap equation with the key input parameters obtained from state of the art \emph{ab initio} calculations, namely the G$_0$W$_0$ method for the quasiparticle band structure and the quantum electrostatic heterostructure (QEH) method for the screened Coulomb interaction\cite{andersen2015}. As such, our work distinguishes itself from previous reports\cite{gupta2020,guo2022}, both in terms of the materials considered and the level of the employed methodology.

\begin{figure*}[ht]
\begin{center}
\includegraphics[width=0.95\textwidth]{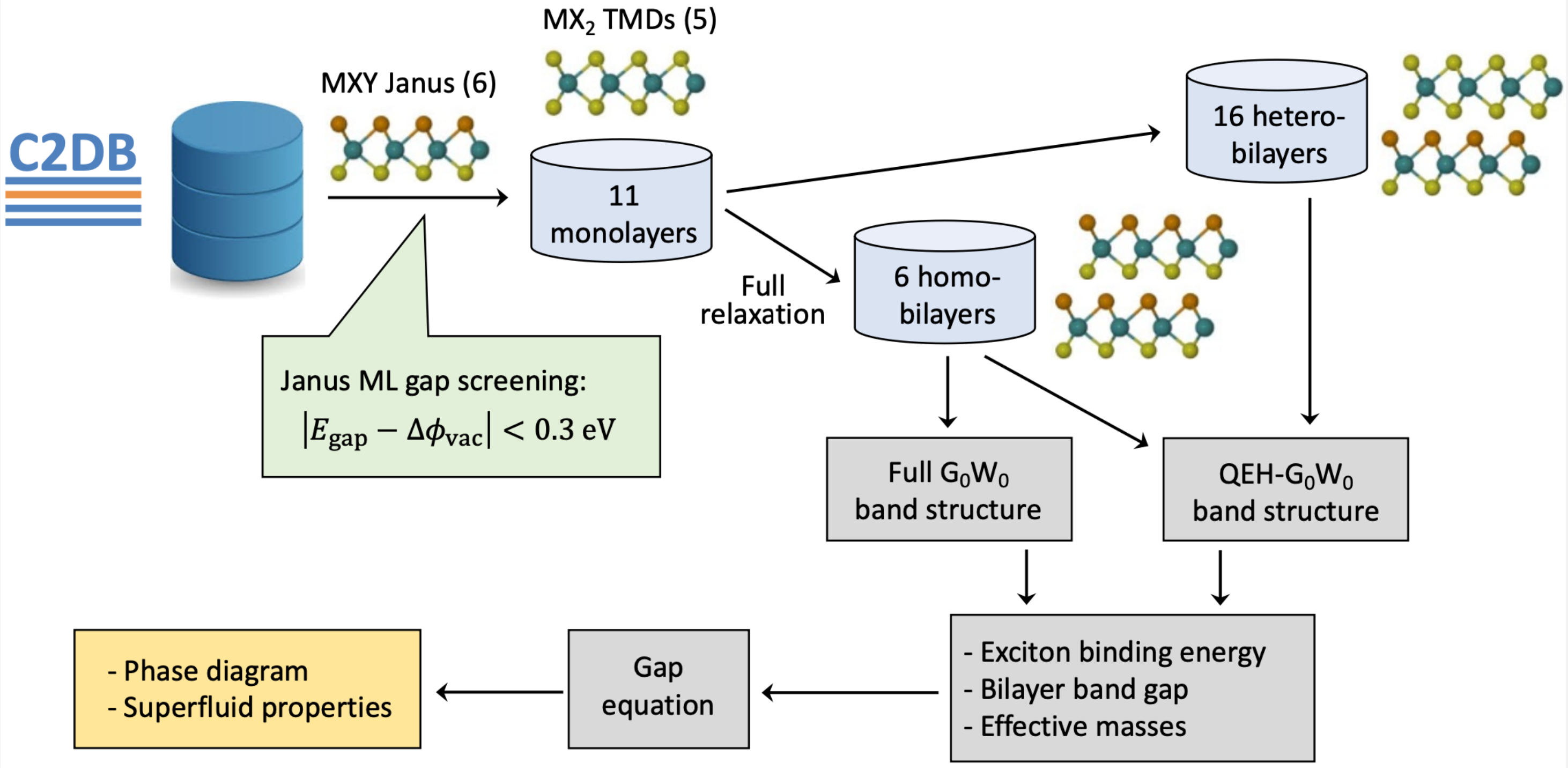}
\caption{Schematic workflow of the calculations. First we select all Janus materials which would provide homo-bilayer gaps between -0.3 to 0.3 eV, according to the Anderson rule \cite{anderson1960}. In addition to these materials we add all experimentally synthesized H-TMDs that can be combined with the other materials to yield hetero-bilayer gaps in the same energy range. We also add MoSSe, which is an experimentally synthesized Janus material but with a too large gap for the initial filtering. This yields a total of 11 materials. For these materials the inter-layer distance and G$_0$W$_0$ band structures are computed for the homo-bilayers using GPAW\cite{mortensen2005,enkovaara2010,Huser2013}, while the band gaps are estimated using the G$_0$W$_0$-QEH\cite{latini2017} model for the hetero-bilayers. Finally exciton binding energies and effective Coulomb interactions are computed using the QEH method. The QEH model also provides the \emph{ab initio} input to the mean-field gap equation that is used to investigate the superfluid properties of the different bilayers.} 
\label{schematic}%
\end{center}
\end{figure*}

The general approach is sketched in Fig. \ref{schematic}. 
We begin by searching the C2DB\cite{c2db1,c2db2,c2db3} for dynamically and thermodynamically stable MXY-Janus materials with a vacuum level difference $\Delta \phi_{\mathrm{vac}}$ close in magnitude to the G$_0$W$_0$ band gap of the monolayer $|E_{\mathrm{gap}}-\Delta \phi_{\mathrm{vac}}| < 0.3$ eV. Furthermore, for computational reasons we concentrate on materials that have valence band maxima (vbm) and conduction band minima (cbm) at high-symmetry points. Out of the initial pool of 106 MXY Janus materials this leaves us with five candidate materials. In addition to these materials we consider transition metal dichalcogenides (TMDs) in the H-phase that have been synthesized in monolayer form, as well as MoSSe which also has been synthesized but has a too small vacuum level difference to qualify in our initial search. We then consider all possible bilayers which can be formed by combining these monolayers, and compute the band alignments using the Anderson rule. Finally, monolayers that do not form any bilayers with a gap between -0.3 and 0.3 eV are filtered out. This provides us with our final set of 11 monolayer materials that we will consider for further analysis. 

\begin{figure*}[ht!]
\begin{center}
\includegraphics[width=0.4\textwidth]{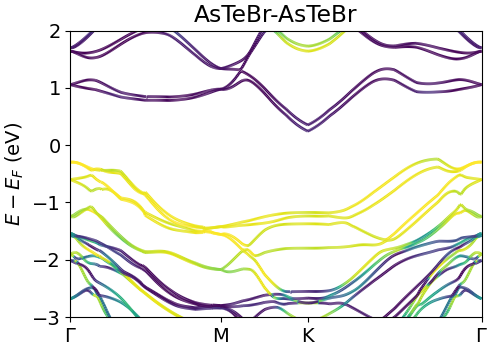}
\includegraphics[width=0.4\textwidth]{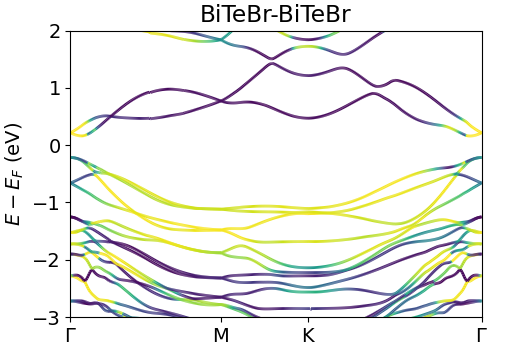}
\includegraphics[width=0.4\textwidth]{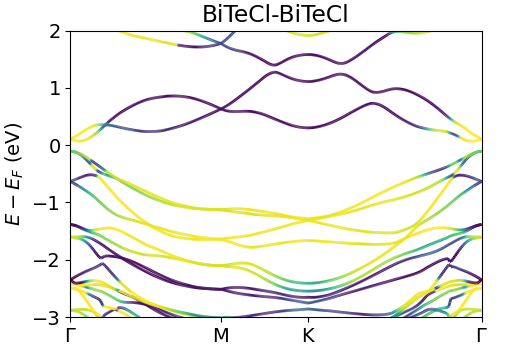}
\includegraphics[width=0.4\textwidth]{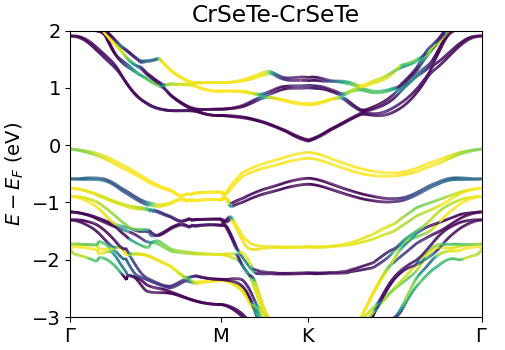}
\includegraphics[width=0.4\textwidth]{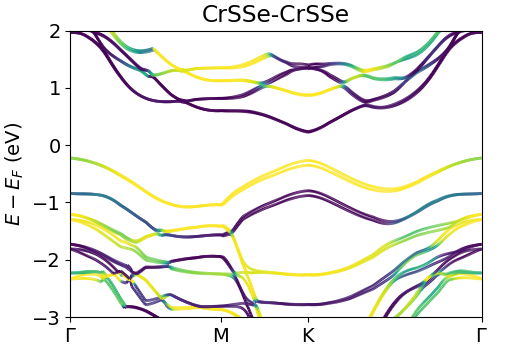}
\includegraphics[width=0.4\textwidth]{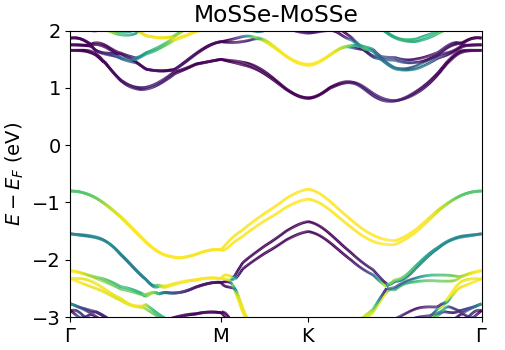}
\includegraphics[width=0.35\textwidth]{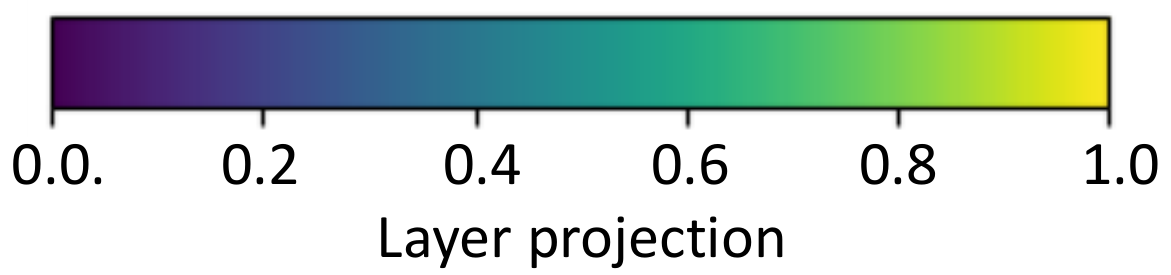}
\caption{Band structures computed with the G$_0$W$_0$ method for Janus homo bilayers. The color bar shows the projection of the PBE+SOC eigenstates onto the bottom layer. Since the PBE+SOC band structure is metallic for BiTeBr and BiTeCl, the layer projections are not accurate for these two materials around the $\Gamma$-point.} 
\label{gw-band}%
\end{center}
\end{figure*}

We proceed to calculate the equilibrium stacking order and interlayer distance for the homobilayers using a $z$-scan approach (see SI). 
For the heterobilayers, the interlayer distances are estimated as the average interlayer distances of the corresponding homobilayers  (we do not perform explicit DFT calculations for the heterobilayers because of the different in in-plane lattice constants, which would lead to very large simulation cells). We construct all possible bilayers and estimate the band gaps from the G$_0$W$_0$+SOC band gaps of the monolayers including an additional screening correction computed with the G$_0$W$_0$-QEH model \cite{winther2017band}:
\begin{equation}\label{eq:gap}
    E_{\mathrm{gap}} = E_{\mathrm{cbm}}^{e} - \phi_{\mathrm{vac}}^{e} - E_{\mathrm{vbm}}^{h} + \phi_{\mathrm{vac}}^{h} - \Delta E^{\mathrm{scr}}. 
\end{equation}
 Here the superscript $e$($h$) denotes the electron (hole) layer and  $E_{\mathrm{cbm/vbm}}$ are the G$_0$W$_0$ band edges of the monolayers relative to the vacuum level (average vacuum level, $\phi_{\mathrm{vac}}^{\mathrm{av}}$,  for Janus monolayers). The terms $\phi_{\mathrm{vac}} = \phi_{\mathrm{vac}}^{\mathrm{av}} - \frac{\phi_{\mathrm{vac}}^{+z}-\phi_{\mathrm{vac}}^{-z}}{2}$, where $\phi_{\mathrm{vac}}^{\pm z}$ are the vacuum levels on the two sides of the sheet, account for the shift in band edges due to the build-in field of Janus monolayers. Finally, $\Delta E^{\mathrm{scr}}$, represents the shifts in the monolayer band edges due to the additional dielectric screening stemming from the other layer. The latter is evaluated using the QEH model\cite{winther2017band}. 
For the six Janus homobilayers, we also perform full G$_0$W$_0$ calculations (Fig. \ref{gw-band}), including spin-orbit coupling. These calculations provide very accurate quasiparticle band gaps for this subset of materials (which is the main focus of the article) and serve as benchmarks for the approximate band gaps obtained from Eq. (\ref{eq:gap}). We note that full G$_0$W$_0$ calculations are very challenging for general heterobilayers due to the different in-plane lattice constants.

\begin{figure*}[ht!]
\begin{center}
\includegraphics[width=0.75\textwidth]{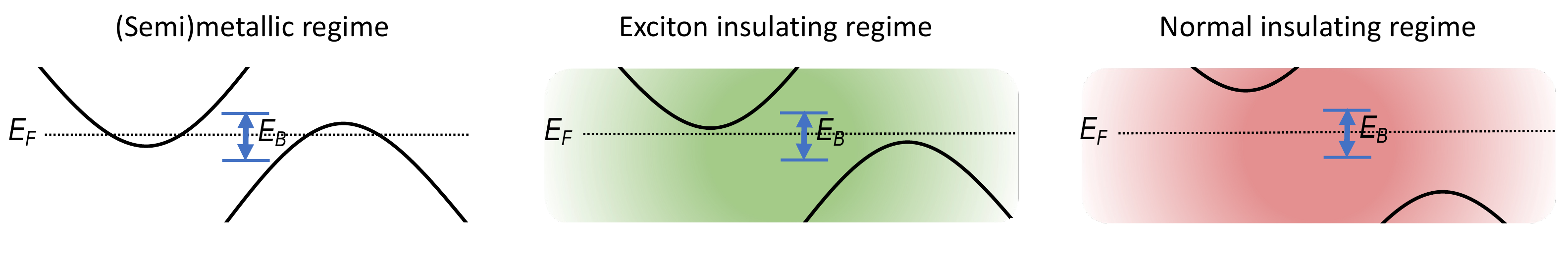}
\includegraphics[width=0.48\textwidth]{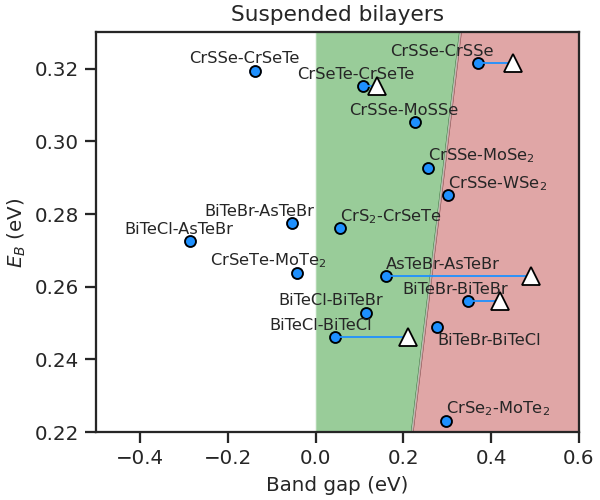}
\includegraphics[width=0.48\textwidth]{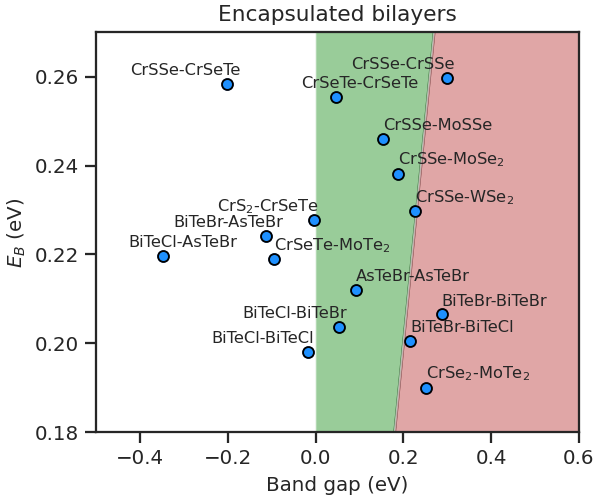}
\caption{Exciton binding energies ($E_B$) as a function of the band gap for Janus homobilayers and heterobilayers. Binding energies for suspended bilayers are shown in the left panel and for bilayers encapsulated in 3 layers of BN on each side in the right panel. The white triangles correspond to QP gaps computed within the G$_0$W$_0$ approximation (for the Janus homobilayers) while the blue circles correspond to QP gaps computed with the G$_0$W$_0$-QEH method. The exciton binding energies are obtained from a Wannier-Mott model with screened e-h interaction calculated with the QEH model. The top panel schematically depicts the relative magnitude of the exciton binding energy and the gap in the three regions.}
\label{exciton-be}%
\end{center}
\end{figure*}

Using effective masses for the relevant bands (cbm in the electron layer and vbm in the hole layer) obtained from the C2DB and averaged over the $x$ and $y$ directions, we compute the interlayer exciton binding energies using a Wannier-Mott model 
\begin{align}
    \left( -\frac{\nabla^2_{2D}}{2\mu} + U^{eh}(\mathbf{r}) \right)F_i(\mathbf{r}) = E_i F_i(\mathbf{r}),
    \label{wanniermott}
\end{align}
where $U^{eh}$ is the screened electron-hole interaction obtained with the QEH model \cite{andersen2015}. Upon comparing the binding energy of the lowest exciton with the band gap, we obtain the final list of 16 bilayers for which the exciton binding energies are larger or only slightly smaller than the QP gap (see Fig. \ref{exciton-be}). These systems are expected to form excitonic insulators and potentially superfluid condensates at low temperatures.


We begin by discussing the G$_0$W$_0$ band structures of the six homobilayers, see Fig. \ref{gw-band}. For these materials the size of the band gap is determined by a delicate competition between spin-orbit coupling (SOC), hybridization and screening. While the PBE band gaps are relatively large, the addition of SOC greatly reduce the band gaps and turn the materials with heavy elements into metals. However, inclusion of correlation effects within the G$_0$W$_0$ self-energy approximation increases the size of the band gap again. For accurate estimations of the gaps for these materials it is therefore essential to include both SOC and correlation effects.  
Monolayer AsTeBr exhibits a small indirect band gap of 0.49 eV while all other materials have direct bandgaps.
The color code indicates the layer projection of the corresponding PBE+SOC Bloch states onto the bottom layer. Most of the materials have bands with well defined layer projections, however, for the Bi-compounds the character of the bands is clearly mixed around the $\Gamma$-point. A large part of this mixing can be attributed the fact that the PBE+SOC band structure is metallic which yields an increased hybridization between the layers in the PBE+SOC band structure compared to the G$_0$W$_0$ band structure.
If the G$_0$W$_0$ wave functions would have been computed self-consistently, as done in quasiparticle self-consistent GW, one would thus expect the hybridization to be reduced and the vbm/cbm states at $\Gamma$ have more distinct layer projections.

The G$_0$W$_0$ quasiparticle corrections are computed on a Monkhorst-Pack (MP) mesh and interpolated to the band structure path in the Brillouin zone. Non-analytic features in the band structure, such as the V-shape of the conduction band in CrSSe-CrSSe are artifacts of the interpolation and not present in the underlying PBE bandstructures. However, since both the $\Gamma$ and $K$ points are included in the MP-mesh the size of the band gaps are accurate and independent of the interpolation.
\\

The exciton binding energy in Eq. (\ref{wanniermott}) is determined by the effective mass and the screened electron-hole Coulomb interaction. Since the dominant contribution to the density response function comes from interband transitions within the individual monolayers, the magnitude of the attractive electron-hole interaction is essentially determined by the monolayer band gap: if the monolayer gap is large, the screening is correspondingly low and the exciton binding energy will be larger. Thus, the reduction of the band gap in the bilayers, which is driven by out-of-plane dipole of the Janus structures, does not enhance the dielectric screening significantly. This makes it possible to construct small band gap materials with large exciton binding energies and is the key advantage of vdW bilayers for realising excitonic ground states.

In the left panel of Fig. \ref{exciton-be} we show the exciton binding energies, computed with the Wannier-Mott model Eq. (\ref{wanniermott}) as a function of the QP gap for all 16 bilayers. One can identify three distinct regions: (1) The (semi)metallic region (white) with a negative QP gap. (2) The excitonic insulating region (green) where the exciton binding energy is larger than the QP gap. (3) The normal insulating region (red) where the gap is larger than the exciton binding energy. One would expect interesting excitonic phases in the white and green region only, since in the red region spontaneous exciton formation is inhibited by the large band gap. All materials in the green region are expected to have an excitonic insulating ground state, while the materials in the white region can be either normal metals or exciton superfluids at low temperatures depending on the competition between metallic screening and exciton condensation. This will be further analysed using the mean field gap equation. 

For the Janus homobilayers we show the band gaps obtained from full G$_0$W$_0$ calculations as white triangles while the G$_0$W$_0$-QEH gaps are shown as blue circles. For vdW heterostructures composed of 2D materials with no out-of-plane dipole, i.e. non-Janus monolayers, 
the Anderson rule for band alignment typically overestimates the band gap due to neglect of hybridisation effects. \cite{winther2017band}. Thus, one would expect the full G$_0$W$_0$ band gaps to be smaller than the G$_0$W$_0$-QEH estimates. Looking at Fig. \ref{exciton-be} it is evident that this simple picture does not hold for most of the Janus homobilayers considered in this work. The effect is most pronounced for AsTeBr where the G$_0$W$_0$-QEH calculation underestimates the full G$_0$W$_0$ gap by more than 0.3 eV. This surprising result can be understood from nontrivial hybridization effects and incomplete dipole band shifts in the Janus bilayers.
 In the G$_0$W$_0$-QEH method one assumes that the electric dipoles of the monolayers are unchanged when the bilayer is formed and hence the vacuum level difference of the bilayer should be twice the vacuum level difference of the monolayers. From Table \ref{tablevacshift} one can see that the vacuum level shift for the six Janus homobilayers ($\Delta \phi_{\mathrm{vac}}^{\mathrm{BL}}$) are in general smaller than twice the monolayer shifts ($\Delta \phi_{\mathrm{vac}}^{\mathrm{ML}}$).
However, this fact alone is not sufficient to explain the large deviations between the G$_0$W$_0$-QEH and full G$_0$W$_0$ calculations for AsTeBr. For this material the vacuum level difference of the monolayer is 1.7 eV, and half the vacuum level difference of the bilayer is 1.4 eV but the shift of the conduction band between the two layers is only 1.16 eV. Thus, the bands are shifted less than half the bilayer vacuum level difference. 
The magnitude of this (incomplete) dipole band shift depend on the stacking, and thus the stacking sequence could potentially be used to fine-tune the gap for individual bilayers. 

To increase the stability of the bilayers it is often advantageous to encapsulate them in a stable material, such as hexagonal boron nitride (hBN). In the right panel we therefore show the same quantities for bilayers encapsulated in three layers of BN on both sides. As expected the addition of the six hBN layers increases the screening in the materials and therefore reduces the effective electron-hole interactions and thus also the exciton binding energies. However, since the QP gap is also reduced, encapsulation does not change any of the qualitative conclusions in this manuscript. It is interesting to note that the reduction is material dependent so that e.g. the relative ordering in binding energy between BiTeBr-AsTeBr and CrS$_2$-CrSeTe is reversed for the encapsulated bilayers. This further highlights the importance of an \emph{ab initio} treatment of the screening in these systems.   

To investigate the superfluid properties, we combine the QEH model with the recently developed exciton DFT method \cite{nilsson2021} to solve a superfluid gap equation with \emph{ab initio} parameters (see supplementary information (SI)). This combination is also interesting from a methods perspective since it shows how the QEH model can be used to compute inputs to general gap-equations and thereby place model studies, such as those in Refs. \citenum{conti2022,pascucci2022,gupta2020,conti2020,conti2020TMD,conti2019,conti2017,perali2013,zarenia2014,lozovik2012,dePalo2002}, in a more stringent \emph{ab initio} framework. 
This \emph{ab initio} downfolding method, as well as the standard model treatment is described in the SI. 

\begin{figure*}[ht]
\begin{center}
\includegraphics[width=0.75\textwidth]{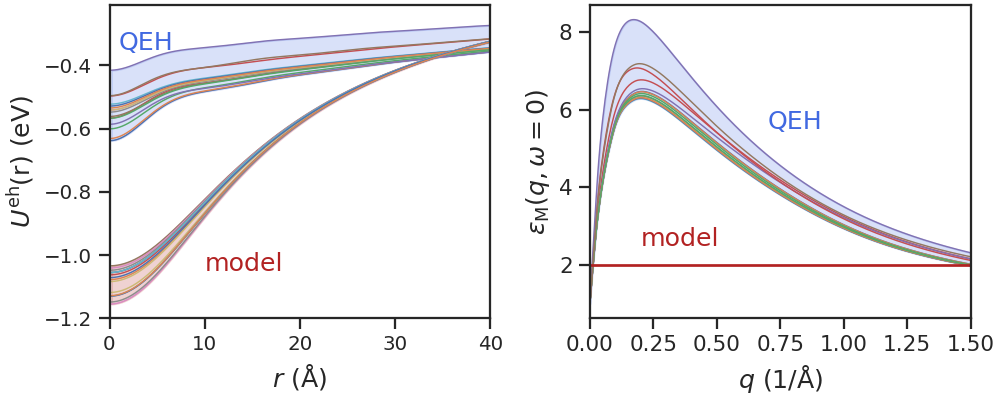}\caption{
Effective electron-hole interaction $U^{eh}(r)$ (left panel) and $q$-dependent static macroscopic dielectric function (right panel). The \emph{ab initio} quantities from the QEH method in blue are compared to the corresponding quantities in the standard model treatment in red (see text and SI). The color of the different lines correspond the different bilayer combinations. The same coloring is used for both the model and QEH results. 
} 
\label{screening}%
\end{center}
\end{figure*}

In Fig. \ref{screening} we compare the effective electron-hole interaction $U^{eh}$ and the macroscopic dielectric function to the corresponding quantities in the standard model treatment. The screening in the model interaction (red lines) is determined by a single parameter, the static dielectric constant, which is taken to have the value two for a wide range of different materials (see the right panel) \cite{conti2022,pascucci2022,gupta2020,conti2020,conti2020TMD,conti2019,conti2017,perali2013,zarenia2014,lozovik2012,dePalo2002}. Thus, the only material dependent quantity that enters the model interaction is the distance between the electron and hole layers. In contrast, the \emph{ab initio} calculated dielectric function (blue lines) has both a substantial $q$ and material dependence, indicating a pronounced spatial and material dependence in the screening which yields a qualitatively different electron-hole interaction (left panel).
 It is not only the depth of the potential well that is different between the \emph{ab initio} and model interaction, but also the relative ordering between the different materials. Therefore a careful treatment of the screening is essential to understand and predict material trends.

\begin{figure*}[ht]
\begin{center}
\includegraphics[width=0.32\textwidth]{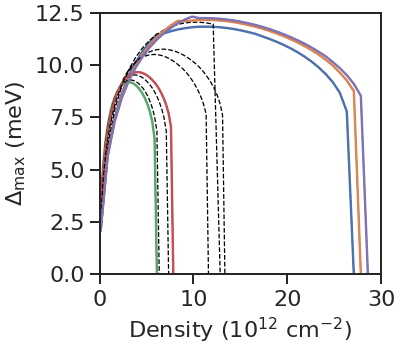}
\includegraphics[width=0.32\textwidth]{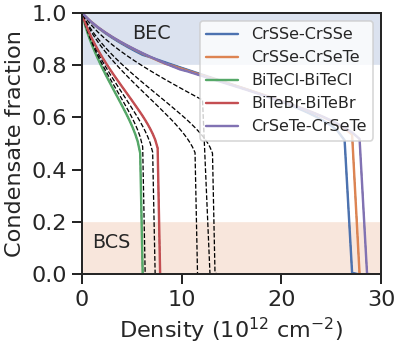}
\includegraphics[width=0.32\textwidth]{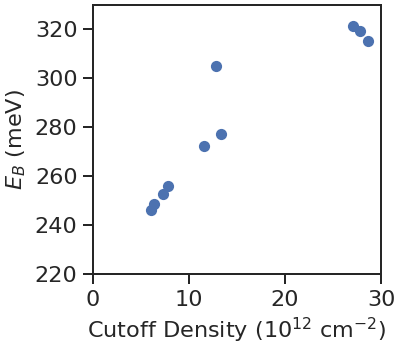}
\caption{Maximum superfluid gap (left) and condensate fraction (middle) as function of density for the different bilayers. For convenience only the most interesting bilayers are labelled, the remaining materials are shown with dashed lines. In the right panel the cutoff-density, i.e. the density where $\Delta$ vanishes, is compared to the exciton binding energy from Eq. \ref{exciton-be}.}
\label{superfluidfig}%
\end{center}
\end{figure*}

We then proceed to solve the mean-field gap equation (see SI) for all bilayers with direct band gaps where CDW reconstructions are not expected to be associated with spontaneous exciton formation, since the excitons do not carry a finite momentum.\cite{Jiang2018} Superfluidity survives at relatively high electron and hole densities (left panel in Fig. \ref{superfluidfig}), up to 30 $\times 10^{12}$ cm$^{-2}$ for the Cr compounds, with the highest cutoff density for bilayer CrSeTe corresponding to a transition temperature $T_{KT} \approx 46$ K. 

The superfluid state is within the mean-field description characterized by the condensate fraction $c = \frac{\sum_k u_k^2 v_k^2}{\sum_k v_k^2}$, where $u_k$ and $v_k$ are the Bogiliubov amplitudes in the mean-field model. For $c>0.8$ the condensate is said to be in the BEC regime of spatially localized electron-hole pairs while for $c<0.2$ it is said to be in the BCS regime. This characterization is, within the mean-field model, approximate but has shown to be useful for a range of systems\cite{Neilson2014,conti2017, conti2019,conti2020, conti2020TMD, nilsson2021, conti2020}.  As can be seen in the middle panel in Fig. \ref{superfluidfig} the gap collapses in the crossover regime, before the systems have reached the BCS regime. 

In this work we use a static mean-field approximation for the self-energy in the gap equation (see SI) and thereby ignore the effect of dynamical screening. In Ref. \citenum{nilsson2021} it was shown that an effective treatment of the frequency dependence beyond the mean-field approximation yield increased cutoff densities and transition temperatures. Similarly, it has been shown that the parabolic approximation to the dispersion tend to underestimate the cutoff densities for TMD heterostructures.\cite{conti2020TMD} Since we employ both these approximations the transition temperatures and cut-off densities in this work should be regarded as a lower limit.

It is interesting to compare the exciton binding energy within the Wannier-Mott model and the superfluid cut-off density within the mean-field gap equation. Since both these quantities in principle are uniquely determined by the effective mass and effective interaction one would naively expect that there would be some correlation between the two quantities. Indeed, as can be seen in the right panel of Fig. \ref{superfluidfig} there is a correlation between high binding energies and high cut-off densities, albeit not very clear. The reason for the fluctuations is twofold. First, contrary to the superfluid properties, the exciton binding energy does not depend on the electron-electron or hole-hole interactions. Secondly, the metallic screening in the doped systems, which determine the fully screened interaction in the gap equation, depend in a non-trivial way on the single-particle dispersion, superfluid gap and effective interactions $U$ through the self-consistency in the gap equation. While the exciton binding energy in the Wannier-Mott model increases for larger effective masses, large effective masses would for a fixed $W^{eh}$ tend to increase the superfluid gap in the gap equation, but for a fixed $\Delta$ tend to increase the screening and thereby reduce $W^{eh}$. Thus it is not sufficient to look at the exciton binding energies in order to determine the superfluid properties of the doped systems.

\begin{table}[ht!]
\centering
\begin{tabular}{lllrrrrrrrr}
\hline
 Material   &   $\Delta \phi_{\mathrm{vac}}^{\mathrm{ML}}$ &   $\Delta \phi_{\mathrm{vac}}^{\mathrm{BL}}$ \\
\hline
 CrSSe      &        -0.826 &           -1.334 \\
 CrSeTe     &        -0.783 &           -1.18  \\
 BiTeCl     &         1.547 &            2.807 \\
 BiTeBr     &         1.204 &            2.229 \\
 AsTeBr     &         1.716 &            2.863 \\
 MoSSe      &        -0.755 &           -1.328 \\
\hline
\end{tabular}
\caption{Comparison of the monolayer and bilayer vacuum level differences for the 6 Janus materials considered in this work.}
\label{tablevacshift}
\end{table}

To summarize, we propose a new platform for excitonic insulators and exciton superfluidity, namely bilayers of 2D Janus materials. In these systems the out of plane dipole yields a reduced gap in the bilayer compared to the monolayers and it is thus possible to form bilayers with a narrow and "spatially indirect" band gap from two monolayers with relatively large intrinsic band gaps. We show that the interlayer exciton binding energies in these systems can exceed the bilayer band gap, and therefore one would expect the materials to have an exciton insulating ground state. 
An analysis of the superfluid properties of the bilayers as a function of electron and hole concentrations shows that superfluid transition temperatures above 50 K are possible. The most promising compounds for exciton superfluidity are bilayers based on the chromium-chalcogenide MXY Janus materials CrSSe and CrSeTe. 

\begin{suppinfo}
The Supplemental information includes all relevant properties of the monolayer building blocks and bilayer structures as well as an account of the z-scan approach used to determine the interlayer distance. We also include a detailed discussion of the gap equation used to model the superfluid properties and a motivation for the \emph{ab initio} downfolding scheme used to determine the effective interaction in the gap equation. Finally, the computational parameters are stated.
\end{suppinfo}

\begin{acknowledgement}
F. N. would like to acknowledge discussions with Chin-Shen Ong at Uppsala University in the initial phases of this project.
The authors acknowledge funding from the European Research Council (ERC) under the European Union’s Horizon 2020 research and innovation program Grant No. 773122 (LIMA) and the Villum Investigator Grant No. 37789 supported by VILLUM FONDEN.
F.N has received funding from the European Union’s Horizon 2020 research and innovation program under the Marie Skłodowska-Curie grant agreement No. 899987. (EuroTechPostdoc2).
\end{acknowledgement}

\bibliography{refs}
\end{document}